
\magnification 1200
\baselineskip=6truemm


\def\plb#1#2#3{{\it Phys. Lett.} {\bf #1B} (19#2) #3}
\def\prd#1#2#3{{\it Phys. Rev.} {\bf D#1} (19#2) #3}

\def\prep#1#2#3{{\it Phys.~Rep.} {\bf C#1} (19#2) #3}


\def\shat{\hat s}
\def\sigmahat{{\hat\sigma}}
\def\lntwoab{\ln\left[{2-\alpha-\beta}\over{2-\alpha+\beta}\right]}
\def\lnab{\ln\left[{\alpha+\beta}\over{\alpha-\beta}\right]}
\def\regu{{m_q^2\over\shat}}

\def\small#1{{\scriptscriptstyle #1}}
\def\Scl{{\cal L}}

\def\sla#1{\raise.15ex\hbox{$/$}\kern-.57em #1}
\def\geapprox{\lower.6ex
              \hbox{\kern+.2em$\buildrel >\over\sim$\kern+.2em}}
\def\leapprox{\lower.6ex
              \hbox{\kern+.2em$\buildrel <\over\sim$\kern+.2em}}
\def\ms{M_{{\scriptscriptstyle S}}}
\def\qs{Q_{{\scriptscriptstyle S}}}
\def\umax{u_{max}}
\def\umin{u_{min}}
\def\lnsu{\ln\left[{\shat+\umax \over \shat+\umin}\right]}
\def\lnu{\ln\left[{\umax\over\umin}\right]}

\def\backscatter{1}
\def\buchwyler{2}
\def\leurer{3}
\def\nadeau{4}
\def\hewpak{5}
\def\ruckl{6}
\def\ginzphoton{7}
\def\effphoton{8}
\def\efffermion{9}

\line{\hfil UdeM-LPN-TH-93-152}
\line{\hfil McGill-93/23}
\line{\hfil hepph-9307324}
\line{\hfil July, 1993}
\null\vskip 3truecm
\centerline{\bf Single Leptoquark Production at $e^+e^-$ and $\gamma\gamma$
Colliders}
\vskip 2truecm
\centerline{G. B\'elanger${}^a$, D. London${}^a$ and H. Nadeau${}^b$}
\vskip .15in
\centerline{\it ${}^a$ Laboratoire de Physique Nucl\'eaire, Universit\'e de
Montr\'eal}
\centerline{\it C.P. 6128, Montr\'eal, Qu\'ebec, CANADA, H3C 3J7.}
\vskip .15in
\centerline{\it ${}^b$ Physics Department, McGill University}
\centerline{\it 3600 University St., Montr\'eal, Qu\'ebec, CANADA, H3A 2T8.}
\vskip 1truecm
\centerline{\it ABSTRACT}
\noindent
We consider single production of leptoquarks (LQ's) at $e^+e^-$ and
$\gamma\gamma$ colliders, for two values of the centre-of-mass energy,
$\sqrt{s}=500$ GeV and 1 TeV. We find that LQ's which couple within the
first generation are observable for LQ masses almost up to the kinematic
limit, both at $e^+e^-$ and $\gamma\gamma$ colliders, for the LQ coupling
strength equal to $\alpha_{em}$. The cross sections for single production
of $2^{nd}$- and $3^{rd}$-generation LQ's at $e^+e^-$ colliders are too
small to be observable. In $\gamma\gamma$ collisions, on the other hand,
$2^{nd}$-generation LQ's with masses much larger than $\sqrt{s}/2$ can be
detected. However, $3^{rd}$-generation LQ's can be seen at $\gamma\gamma$
colliders only for masses at most $\sim\sqrt{s}/2$, making their
observation more probable via the pair production mechanism.

\vfill\eject


One of the more interesting environments in which to study physics beyond
the standard model (SM) is at a high-energy linear $e^+e^-$ collider. Not
only are $e^+e^-$ collisions clean, but it will likely be possible to
adjust the centre-of-mass energy. Furthermore, it has been suggested that,
by using backscattered laser beams, an $e^+e^-$ machine can be converted
into an $e\gamma$ or $\gamma\gamma$ collider [\backscatter]. This is
particularly exciting, since these different modes may be quite useful for
looking for new physics.

Leptoquarks (LQ's), which are absent in the SM but predicted by many of its
extensions, are one example of the new physics which can be studied at such
machines. These particles, which can have electromagnetic charge
$Q_{em}=-1/3$, $-2/3$, $-4/3$ or $-5/3$, would decay into a lepton and a
quark or antiquark, so the signal would be quite striking. In principle,
LQ's couple to fermions of either helicity. In general, leptoquarks can
have spin 0 or 1, but here we concentrate only on scalar LQ's.

Various processes constrain the strength and nature of the LQ couplings to
fermions. For example, for LQ's of charge $-1/3$ which couple to both
$e^-u$ and $\nu_e d$, rare $\pi$ and $K$ decays constrain the couplings to
be chiral [\buchwyler]. That is, LQ's must couple only to left-handed (LH)
or right-handed (RH) quarks, but not both. For these same LQ's, bounds from
weak universality require that the LH couplings be at most about 10\% of
electromagnetic strength. However, these limits need not necessarily apply
to leptoquarks of other charges.

One of the most stringent constraints on LQ couplings comes from the
absence of low-energy flavour-changing neutral currents (FCNC's). In order
to avoid FCNC's, one typically requires the LQ's to couple within a single
generation only. However, M. Leurer [\leurer] has recently pointed out that
this requirement is in fact impossible to meet in general. Due to
Cabibbo-Kobayashi-Maskawa mixing in the left-handed quark sector, one
cannot simultaneously diagonalize the couplings of the LQ in both the
up-quark and down-quark sector. Thus, if one tries to evade constraints
from FCNC's in the down-quark sector, such as $K^0$-${\overline{K^0}}$ and
$B^0$-${\overline{B^0}}$ mixing, by diagonalizing the LH leptoquark
couplings, $D^0$-${\overline{D^0}}$ mixing will then put very strong limits
on the masses and couplings of left-handed LQ's. There are no similar
constraints for the right-handed LQ's.

Of course, this should not discourage experimentalists from looking for
left-handed LQ's. After all, it is possible that there are other new
particles whose effects in low-energy processes would cancel those due to
leptoquarks. Thus, if a left-handed LQ were discovered, in fact {\it two}
types of physics beyond the SM would have been found: the leptoquark
itself, and the new physics responsible for the cancellations! This
possibility is not totally fantastic, since models which include LQ's will
typically also contain other new particles (scalars, gauge bosons, etc.).

In a previous paper [\nadeau], two of us investigated the production of
scalar  leptoquarks at $e\gamma$ colliders at two values of the
centre-of-mass energy, $\sqrt{s}=500$ GeV and 1 TeV. We showed that LQ's
with masses essentially up to the kinematic limit could be seen, even for
couplings as weak as $O(10^{-3})$-$O(10^{-2})\alpha_{em}$. In this paper we
continue the investigation of single leptoquark production at both $e^+e^-$
and $\gamma\gamma$ colliders, again taking $\sqrt{s}=500$ GeV and 1 TeV.
The $e^+e^-$ case was studied some time ago by Hewett and Pakvasa
[\hewpak], but only for charge $-1/3$ LQ's. Here we do a more complete
analysis. We do not, however, agree with their results.

The most general, model independant Lagrangian with $SU(3)\times
SU(2)\times U(1)$ invariant couplings of the scalar leptoquarks and
conservation of the baryon and lepton numbers [\ruckl] can be separated
into two pieces:
$$ \eqalign{
\Scl_L = &~ g_{1\small{L}}{\overline{q^c_\small{L}}}i\tau_2 l_\small{L} S_1
+ g_{3\small{L}}{\overline{q_\small{L}^c}}i\tau_2 \tau^i l_\small{L} S_3^i
+ h_{2\small{L}} {\overline q}_\small{L} i \tau_2 e_\small{R}R_2^\prime~,
\cr
\Scl_R = &~g_{1\small{R}}{\overline{u_\small{R}^c}}e_\small{R} S_1^\prime
+ {\tilde g}_{1\small{R}}{\overline{d_\small{R}^c}}e_\small{R} {\tilde S}_1
+ h_{2\small{R}} {\overline u}_\small{R}l_\small{L} R_2
+ {\tilde h}_{2\small{R}} {\overline d}_\small{R} l_\small{L}{\tilde R}_2~.
\cr} \eqno(1)
$$
The LH quarks and leptons appear in the standard SU(2) doublets
$q_\small{L}$ and $l_\small{L}$, and the superscript $c$ denotes charge
conjugation. In the above equations, following Leurer [\leurer], we have
defined the `handedness' of the leptoquarks according to the helicity of
the quark/antiquark to which they couple\footnote*{Note that this differs
from the conventions of Ref.~\nadeau, in which the handedness of the LQ is
defined by the helicity of the lepton.}. That is, the LQ's in $\Scl_L$ and
$\Scl_R$ are left-handed and right-handed, respectively. From the above, we
see that the LH leptoquarks transform as either a singlet, doublet or
triplet of $SU(2)_W$, while those coupling to RH quarks are singlets or
doublets. The $R$ and $S$ leptoquarks carry fermion number 0 and 2
respectively, with their subscript indicating the $SU(2)_W$ multiplet to
which they belong.

Despite the rather complicated notation in Eq.~1, for our purposes the only
important properties of the leptoquark are its charge and its handedness.
It is straightforward to verify that, for each of the four possible
electromagnetic charges, there exists a LH and a RH leptoquark. There is
one other important point -- in all processes of interest in this analysis,
only the couplings of the LQ to the charged lepton will enter. We can
therefore take the LQ couplings to be generation-diagonal. There is no
conflict with Leurer's result -- there may indeed be a
LQ-neutrino-quark/antiquark coupling which is not generation-diagonal, but
this is unimportant here. In summary, then, the leptoquark is defined by
its charge, by its handedness, and by the generation of the particles to
which it couples. In this paper we will use the symbol $S$ to denote a
leptoquark, while $q$ will refer to either a quark or an antiquark.

One advantage of the $\gamma\gamma\to \ell q S$ process is that it allows
for the production of leptoquarks of each of the three generations
[\ginzphoton]. As we will see, although $2^{nd}$- and $3^{rd}$-generation
LQ's can indeed be produced in $e^+e^-$ collisions, the cross sections are
much too small to be observed.

Let us first focus on single LQ production in $e^+e^-$ collisions. The
diagrams which give rise to this are shown in Fig.~1. Although the large
number of diagrams may seem daunting, most of these can be neglected. There
is a relatively simple way to ascertain which are important and which are
not. Consider the diagrams of Fig.~1a in which a photon is exchanged. When
the virtual photon is aligned with the positron beam direction, the
amplitude diverges. This divergence is regulated by the small mass of the
positron, giving rise to a logarithmic enhancement of about 30 in the total
cross section. In the following, we will refer to such enhancements as
`large logs'. A quick way to spot diagrams which have large
logs is to look for vertices involving three massless (or nearly massless)
particles with at least one $t$- or $u$-channel propagator. Application of
this rule reveals that none of the diagrams in Figs.~1b or 1c have large
logs, so these diagrams can be neglected, and similarly for those diagrams
in Fig.~1a in which a $Z$ is exchanged. In fact, with this rule one expects
that the second diagram of Fig.~1a with photon exchange should have an
additional large log due to the quark propagator. We will see below that
this is indeed the case.

The presence of these divergences indicates that most of the cross section
comes from a few directions in phase space. This makes it very difficult to
use conventional Monte Carlo methods for computing phase space integrals.
A much simpler way to evaluate the diagrams of Fig.~1a in which a photon is
exchanged is to use the effective photon approximation [\effphoton]:
$$
\sigma(s) = \int_{s_{th}/s}^1 d\tau f_\gamma(\tau) \sigmahat(\tau s),
\eqno(2)
$$
in which $\sigma(s)$ is the cross section for the process $e^+e^- \to e^+ q
S$ at centre-of-mass energy $s$, and $\sigmahat(\tau s)$ is the cross
section for the sub-process $\gamma e^-\to q S$ with centre-of-mass energy
$\shat = \tau s$. The minimum $\shat$ required ($s_{th}$) is $(\ms +
m_q)^2$. The photon distribution function $f_{\gamma}(\tau)$ is
[\effphoton]:
$$
f_\gamma(\tau) = {\alpha\over 2\pi} \left\{
{ \left[1+(1-\tau)^2 \right] \over \tau}
\ln \left[{s\over 4m_e^2}{(1-2\tau+\tau^2)\over (1-\tau+\tau^2/4)}\right]
+ \tau \ln \left( {2-\tau\over\tau} \right)
+ {2(\tau-1)\over \tau} \right\}~. \eqno(3)
$$
As expected, $f_\gamma(\tau)$ contains a large log. Note that the more
common form of this function,
$$
f_{\gamma}(\tau) = \left({\alpha\over 2\pi} \ln{s\over 4m_e^2}\right)
{1+(1-\tau)^2 \over \tau}~, \eqno(4)
$$
is quite adequate when $\ms$ is relatively small compared to $\sqrt{s}$.
However, for large $\ms$ the full form (Eq.~3) must be used.

In order to use Eq.~2, we must evaluate the cross section for $\gamma
e^-\to q S$. The diagrams describing this process are shown in Fig.~2. The
key point which must be addressed is that, in the limit in which the quark
mass is neglected, the second diagram diverges. This corresponds to the
situation in which the photon and the quark are aligned. One way to deal
with this is to impose a $p_{\small{T}}$ cut on the quark jet. This is the
method used by Hewett and Pakvasa [\hewpak]. The problem with this solution
is that, because the large logs are due to that region of phase space in
which the entire event is collinear, one loses a significant fraction of
the total cross section. An alternative procedure, which is the one we
advocate, is to use the nonzero quark mass as a regulator. As we will see,
this results in an enormous enhancement of the total cross section compared
to the $p_{\small{T}}$ cut. Experimentally, the situation is that the
entire event goes down the beam pipe. However, the leptoquark will then
decay into a jet and a lepton, giving a signal in the detector which is
unmistakable: $\gamma e^-~(or~e^+e^-)\to e^- + jet$! For comparison, we
will present both methods of regulating the divergence.

We first consider calculating the diagrams of Fig.~2 neglecting the lepton
mass, but keeping a nonzero mass for the quark, $m_q$. For all leptoquarks
we will use the generic Yukawa coupling constant $g$, with the
understanding that the coupling could depend on the masses involved and
might vary from one generation to the other. We parametrize the strength of
the LQ coupling by comparing it to the electromagnetic interaction, i.e.,
$g^2=4\pi k\alpha_{em}$, and allowing $k$ to vary. Denoting the charge of
the leptoquark by $\qs$, the full expression for $\sigmahat(\shat)$ is then
found to be
$$\eqalign{
\sigmahat(\shat) = & {\pi k \alpha_{em}^2 \beta \over 2 \shat} \left[
{\alpha\over 2} + {\qs \over 2}
\left\{ 4 - 2\alpha + {4\over\beta} \left( 1-\alpha +\regu\right)\lntwoab
\right\} \right. \cr
& + {\qs^2\over 2} \left\{ 12 - 10\alpha + {4\over\beta}
\left( (\alpha- 3) (\alpha-1)+ \regu \right) \lntwoab \right\} \cr
& - {\qs+1\over 2} \left\{ 4-2\alpha+ {4\over \beta}\regu\lnab \right\} \cr
& + {(\qs+1)^2\over 2} \left\{ 8-10\alpha + {1\over\beta}
\left(2- 4\alpha+4\alpha^2+4\regu\right)\lnab \right\} \cr
& + \qs(\qs+1) \left\{ -6+6\alpha+{2\over\beta}\regu (1-2\alpha)\lnab
\right. \cr
& \left. \left.~~~~~~~~~~~~~~~~~~~~~~~~-{2\over\beta}(3-2\alpha)\left(
\regu +1-\alpha\right) \lntwoab\right\} \right], \cr} \eqno(5)
$$
in which
$$
\alpha\equiv 1 - {(\ms^2 - m_q^2)\over\shat} \eqno(6)
$$
and
$$
\beta \equiv\left( 1 - 2{(\ms^2+m_q^2)\over\shat}
+ {(\ms^2-m_q^2)^2\over\shat^2} \right)^{1\over 2}~. \eqno(7)
$$
One important point to notice is that Eq.~5 is independent of the
handedness of the LQ. In other words, the cross section for the subprocess
$\gamma e^-\to q S$ is the same for both LH and RH leptoquarks of charge
$\qs$.

The cross section for single leptoquark production in the process
$e^+e^-\to e^+ q S$ can now be calculated using the effective photon
approximation by substituting Eq.~5 into Eq.~2 and numerically computing
the integral. Our results for $\sqrt{s} = 500$ GeV and 1 TeV appear in
Figs.~3a and 3b, where we have taken the LQ coupling strength to be equal
to that of the electromagnetic interaction, i.e.~$k=1$, and have set
$m_q=7$ MeV (this corresponds roughly to either a $d$-quark or a $u$-quark
mass). Assuming the integrated luminosity at a high-energy $e^+e^-$
collider to be 10 ${\rm fb}^{-1}$ at 500 GeV, and 60 ${\rm fb}^{-1}$ at 1
TeV, and requiring 25 events for discovery, one can see that LQ's almost up
to the kinematic limit can be seen in $e^+e^-$ colliders. More precisely,
LQ's with $\qs=-1/3$ and $-5/3$ are observable if $\ms\le 475$ GeV (960
GeV) at $\sqrt{s}=500$ GeV (1 TeV), and those with $\qs=-2/3$ and $-4/3$
can be seen for $\ms\le 420$ GeV (870 GeV) at $\sqrt{s}=500$ GeV (1 TeV).
These numbers are given explicitly in Table 1, where they can be compared
with the prospects at $\gamma\gamma$ colliders, which we will discuss later
in the paper. Since the cross section is linear in $k$, it is
straightforward to scale the results shown in Fig.~3 to other values of
$k$, if desired. It should also be stressed that, because we have used an
approximation in the calculation, there is some uncertainty in the above
numbers, perhaps as much as 5\% [\efffermion].

As noted above, most of the cross section comes from that region of phase
space in which the entire event goes down the beam pipe. Since the LQ
decays to $\ell+jet$, there will be essentially no background from SM
processes. Even for those events in which other particles are seen in the
detector, there will be a sharp invariant mass peak in $M_{\ell+jet}$ at
$\ms$, which is not present in SM decays.

One interesting feature of Fig.~3 is that the cross sections for the LQ's
with $\qs=-1/3$ and $\qs=-5/3$ are almost equal, and similarly for those
LQ's with $\qs=-2/3$ and $-4/3$. This reflects the dominance of the second
diagram in Fig.~2, since it has an extra large log compared to the other
two. Since the amplitude for this diagram is proportional to the quark
charge, $Q_q = -(\qs+1)$, the most important term in $\sigmahat(\shat)$ is
the one whose coefficient is $(\qs+1)^2$. From this it follows that, to a
very good approximation, the cross sections for LQ's with $\qs=-1/3$ and
$-5/3$ should be equal. Similarly, LQ's with $\qs=-2/3$ and $-4/3$ are
expected to have equal cross sections, and these should be a factor of 4
smaller than the cross section for the LQ with $\qs=-1/3$. These
expectations are born out in Fig.~3.

We have also calculated the diagrams in Fig.~2 by imposing a
$p_{\small{T}}$ cut on the quark jet. In this case, the cross section for
$\gamma e^- \to q S$ takes the form
$$\eqalign{
\sigmahat(\shat) = & {\pi k \alpha_{em}^2 \over 2 \shat^2} \left[
- {1\over 2\shat} \left(\umax^2-\umin^2\right) \right. \cr
& + {2(\qs+1)\over\shat} \left\{ {1\over 2}\left(\umax^2-\umin^2\right)
- \ms^2\left(\umax-\umin\right) \right\} \cr
& -{2\qs\over\shat} \left\{ {1\over 2}\left(\umax^2-\umin^2\right)
- \ms^2\left(\umax-\umin\right) + \shat\ms^2\lnsu \right\} \cr
& -(\qs+1)^2 \left\{ {1\over\shat}\left(\umax^2-\umin^2\right)
+ \left(2-{4\ms^2\over\shat}\right) \left(\umax-\umin\right) \right. \cr
& \left. ~~~~~~~~~~~~~~~~~~~~ + \left(-2 \ms^2 + \shat +
{2(\ms^2)^2\over\shat}\right) \lnu \right\} \cr
& + {4\qs(\qs+1)\over\shat} \left\{ {1\over 2}\left(\umax^2-\umin^2\right)
+ \left(-2\ms^2 + {\shat\over 2}\right) \left(\umax-\umin\right) \right.
\cr
& \left. ~~~~~~~~~~~~~~~~~~~~~~~~+ \ms^2 \left( \ms^2+{\shat\over 2} \right)
\lnsu \right\} \cr
& - {2\qs^2\over\shat} \left\{ {1\over 2}\left(\umax^2-\umin^2\right)
- 2\ms^2\left(\umax-\umin\right) \right. \cr
& \left. \left. ~~~~~~~~~~
+ \shat (\ms^2)^2 \left[ {1\over \shat+\umax} - {1\over \shat+\umin} \right]
+ \ms^2 \left(\ms^2 + 2 \shat\right) \lnsu
\right\} \right]. \cr} \eqno(8)
$$
Here,
$$
\umax = -{1\over 2}\shat\,\beta' \left(1-c_{max}\right)~,~~~~~
\umin = -{1\over 2}\shat\,\beta' \left(1+c_{max}\right)~, \eqno(9)
$$
with
$$
\beta' = 1 - {\ms^2\over\shat}~,~~~~~
c_{max}=\sqrt{1-{2 p_{\small{T}cut} \over \sqrt{\shat}\,\beta'}}~, \eqno(10)
$$
in which $p_{\small{T}cut}$ is the imposed $p_{\small{T}}$ cut which we
take to be 10 GeV.

We now calculate as before the cross section for single leptoquark
production in the process $e^+e^-\to e^+ q S$ using the $\sigmahat(\shat)$
given in Eq.~8. In order to compare with the results of Hewett and Pakvasa
[\hewpak], we consider $\qs=-1/3$ LQ's only, use $\sqrt{s}=1$ TeV, and take
$k=2$ (due to a difference of $\sqrt{2}$ in the definition of the LQ
coupling, this corresponds to $k=1$ in the notation of Ref.~\hewpak.). The
result is shown in Fig.~4. Our results are significantly smaller than those
obtained in Ref.~\hewpak, by roughly an order of magnitude for all values of
$\ms$. The difference might perhaps be due to the form taken for the photon
distribution function in the effective photon approximation (Eq.~3).

What is certain is that one gains a significant amount in the total cross
section for single leptoquark production by {\it not} imposing a
$p_{\small{T}}$ cut, but rather using the nonzero quark mass to regulate
the collinear divergence. For example, compare Fig.~3b with Fig.~4
(remembering to rescale the numbers one reads off of Fig.~4 by a factor of
2 in order to correspond to our $k=1$). A LQ of mass 800 GeV has a cross
section of about 4 fb if the quark mass method is used. On the other hand,
using the $p_{\small{T}}$ cut, an 800 GeV LQ would have a cross section of
0.4 fb, which is considerably smaller. For the rest of the paper, we will
restrict ourselves to evaluating the large logs using a nonzero quark mass.

Since we have assumed that each LQ couples generation-diagonally, the
production cross sections shown in Fig.~3 hold only for first generation
leptoquarks. The only way to produce single $2^{nd}$- or $3^{rd}$
generation LQ's in $e^+e^-$ collisions is through the graphs of Fig.~1c.
Note, however, that there are no large logs in these diagrams, so we expect
the cross sections to be smaller than those of Fig.~3 by at least two to
three orders of magnitude. We have calculated these diagrams explicitly,
and we find that indeed the single LQ production cross sections are
typically $O(10^{-3})$ fb. Thus, there is no hope for seeing single
$2^{nd}$- or $3^{rd}$ generation LQ's in $e^+e^-$ collisions\footnote*{Note
that $2^{nd}$- and $3^{rd}$-generation LQ's could be seen if they were pair
produced in $e^+e^-$ collisions via s-channel $\gamma$- or $Z$-exchange. Of
course, this is only possible for $\ms\le\sqrt{s}/2$.}. It is, however,
possible to observe such LQ's in $\gamma\gamma$ collisions, and we now turn
to a study of such processes.

For the process $\gamma\gamma \to \ell^+ q S$, there are six diagrams,
shown in Fig.~5. In fact, there are really twelve diagrams, since each
final  state must be symmetrized with respect to the initial photons.
Again, it is not necessary to calculate all the graphs -- some can be
neglected. Using the large log counting rules introduced earlier, one finds
that, for the case in which the quark mass is small, the diagram in Fig.~5a
contains two large logs, the four diagrams of Figs.~5b and 5c each have one
large log, while the diagram in Fig.~5d has none. Our experience with
single LQ production in $e^+e^-$ collisions tells us that the graph in
Fig.~5a will essentially completely dominate in this case. For
$3^{rd}$-generation LQ's, the mass of the top quark ($\sim 150$ GeV) can no
longer be considered small compared to the centre-of-mass energy. In this
case, the large log counting changes -- the graphs in Figs.~5a and 5b each
have one large log, while those in Figs.~5c and 5d have none. Therefore,
for LQ's of all generations, it is an excellent approximation to keep only
the diagrams in Figs.~5a and 5b in the calculation of the total cross
section, and this is what we will do.

Note that, although we must include the interference among the three
diagrams of Figs.~5a and 5b in our calculation, we may ignore the
interference between this set of three graphs and the set which is
symmetrized with respect to the initial photons. This can be seen
intuitively as follows. The divergences in one set of graphs, which give
rise to the large logs, occur when the virtual lepton goes in the direction
of one of the initial photons, while the divergences of other set of graphs
are present in that region of phase space in which the virtual lepton
aligns itself with the other photon. Therefore, when these two sets of
graphs interfere, there are no divergences, and hence no large logs. We
have verified this intuitive picture analytically, and find that indeed
there are no large logs coming from the interference of the two sets of
diagrams. Thus, it is only necessary to evaluate the contribution of the
graphs of Figs.~5a and 5b to the process $\gamma\gamma \to \ell^+ q S$, and
then to include a factor of two to take into account the symmetrized set of
diagrams.

The easiest way to calculate the diagrams in Figs.~5a and 5b is to use a
technique similar to that used in the computation of $e^+e^- \to e^+ q S$,
namely the effective fermion approximation. That is,
$$
\sigma(s) = \int_{s_{th}/s}^1 d\tau f_\ell(\tau) \sigmahat(\tau s),
\eqno(11)
$$
in which $\sigma(s)$ is the cross section for $\gamma\gamma \to \ell^+ q S$
at centre-of-mass energy $s$, and $\sigmahat(\tau s)$ is the cross section
for the sub-process  $\gamma \ell^-\to q S$ with centre-of-mass energy
$\shat = \tau s$, just as in the effective photon approximation. The lepton
distribution function is [\efffermion]
$$
f_\ell(\tau) = {\alpha\over 2\pi} \left\{ \left[\tau^2+(1-\tau)^2 \right]
\ln \left[{s\over 4m_\ell^2} (1-\tau)^2 \right] + 2\tau(1-\tau) \right\}~.
\eqno(12)
$$
Again, as expected, this function contains a large log. Although
$f_\ell(\tau)$ somewhat resembles the photon distribution function
$f_\gamma(\tau)$ (Eq.~3), there is one important difference. Due to the
absence of a factor of $\tau$ in the denominator, $f_\ell(\tau)$ is much
smoother than $f_\gamma(\tau)$. Thus we do not expect the cross section for
$\gamma\gamma \to \ell^+ q S$ to be as strong a function of $\ms$ as we
found in $e^+e^- \to e^+ q S$.

The cross section for the subprocess $\gamma \ell^-\to q S$ has already
been calculated (Eqs.~5-7), so we can simply perform the numerical
integration of Eq.~11 for all three generations. We have taken $m_e=0.5$
MeV and $m_u=m_d=7$ MeV for the first generation, $m_\mu=100$ MeV,
$m_s=150$ MeV and $m_c=1.5$ GeV for the second, and $m_\tau=1.5$ GeV,
$m_b=5$ GeV and $m_t=150$ GeV for the third.  We remind the reader that the
LQ's of charge $-1/3$ and $-5/3$ couple to up-type quarks, while the
$Q_{em}=-2/3$ and $-4/3$ LQ's couple to down-type quarks. In computing the
cross section, we have included the factor two to take into account the
symmetrized set of diagrams.

The results are shown in Fig.~6 for the three generations, for
$\sqrt{s}=500$ GeV and 1 TeV, for $k=1$. Before describing the results, let
us note some general features. For the $1^{st}$- and $2^{nd}$-generation
LQ's, the similarity of the curves for LQ's of $Q_{em}=-1/3$ and $-5/3$,
and for LQ's with charges $-2/3$ and $-4/3$, again reflects the dominance
of the diagram in Fig.~5a (two large logs). Also, as expected, the cross
sections are in general less strongly dependent on the LQ mass than was
found in $e^+e^-$ colliders. Finally, the cross sections for
$3^{rd}$-generation LQ's are significantly smaller than for those coupling
within the $1^{st}$- and the $2^{nd}$ generations. This reflects the fact
that there is really one less large log in the cross section for
$3^{rd}$-generation LQ's.

The figure of merit in Fig.~6 is the largest LQ mass observable for each of
the three generations. The question is, which has the better prospects for
LQ detection, the single LQ production mode, or the pair production mode
(in which LQ's of mass $\ms\le\sqrt{s}/2$ can be seen)? Looking at Fig.~6,
if LQ's of mass greater than $\sqrt{s}/2$ can be seen, then it is better to
try to detect leptoquarks in the single LQ production mode. However, if the
maximum LQ mass which can be observed is less than $\sqrt{s}/2$, then pair
production is more promising. In Table 1 we display $(\ms)_{max}$ for all
four LQ charges and for all three generations.

{}From Figs.~6a and 6b and Table 1, we see that, as in $e^+e^-$ colliders,
$1^{st}$-generation LQ's with masses almost up to the kinematic limit can
be seen in $\gamma\gamma \to \ell^+ q S$. As before, we have assumed an
integrated luminosity of 10 ${\rm fb}^{-1}$ at 500 GeV, and 60 ${\rm
fb}^{-1}$ at 1 TeV, and assumed a discovery signal of 25 events. And again,
there is perhaps a 5\% uncertainty in these numbers due to the
approximations used [\efffermion].

The situation is similar, though not quite as promising, for
$2^{nd}$-generation LQ's (Figs.~6c and 6d, Table 1). At $\sqrt{s}=500$ GeV,
the maximum mass allowed for observing a LQ is about 300-400 GeV, depending
on the LQ charge, while at 1 TeV, it is 700-900 GeV. We remind the reader
that we have used the $c$-quark mass in the cross sections for LQ's with
charge $-1/3$ and $-5/3$, and the $s$-quark mass for LQ's with
$Q_{em}=-2/3$ and $-4/3$.

Things are very different for $3^{rd}$-generation LQ's (Figs.~6e and 6f,
Table 1). At $\sqrt{s}=500$ GeV, the maximum mass is 100-200 GeV, except
for the charge $-1/3$ LQ, which is not observable at all. At 1 TeV,
$(\ms)_{max}$ is 180-540 GeV. For $3^{rd}$-generation leptoquarks, then, it
is almost always better to look for pair production in $e^+e^-$ or
$\gamma\gamma$ collisions. It should be emphasized, however, that these
cross sections have been calculated for LQ coupling strengths $k=1$. If the
LQ couplings were proportional to masses, then for those LQ's which couple
to the $t$-quark one might conceivably have $k$ larger than one, and the
cross sections would increase accordingly. Of course, if $k$ were much
larger than one, then at some point this perturbative analysis would break
down.

An important point to remember is that, in $\gamma\gamma$ colliders created
by the backscattering of laser light, the photon beams are not
monochromatic. For a complete calculation it would be necessary to fold in
the energy spectrum of the initial photons. Typically the highest energy
photons would have about 80\% of the energy of the parent electron machine
so that the limits given in Table 1 would be scaled accordingly.

In conclusion, we have calculated the cross sections for single leptoquark
production at high-energy $e^+e^-$ and $\gamma\gamma$ colliders of
$\sqrt{s}=500$ GeV and 1 TeV. For LQ's coupling within each of the
$1^{st}$, $2^{nd}$ and $3^{rd}$ generations, we have considered the four LQ
charges $\qs=-1/3,-2/3,-4/3,-5/3$. Our results are independent of whether
the LQ couples to left- or right-handed quarks/antiquarks.

For the process $e^+e^- \to e^+ q S$ we have utilized the effective photon
approximation, while for $\gamma\gamma\to \ell^+ q S$ the effective fermion
approximation was used. For each of these methods it was necessary to
calculate the cross section for the subprocess $\gamma \ell^-\to q S$. We
have shown that using a $p_{\small{T}}$ cut to regulate the collinear
divergence in this process is in fact not a very good procedure -- one
loses too much of the total cross section. It is better to use the nonzero
quark mass as a regulator. In this case the bulk of the cross section comes
from that region of phase space in which the entire event goes down the
beam pipe. When the LQ decays, this results in an unmistakable signal in
the detector: $e^+e^- (or~\gamma\gamma) \to \ell^- + jet$. This is
virtually background-free.

We have found that $1^{st}$-generation LQ's of any charge can be observed
almost up to the kinematic limit in both $e^+e^-$ and $\gamma\gamma$
colliders ($\sqrt{s}=500$ GeV or 1 TeV), for LQ coupling strengths equal to
that of the electromagnetic interaction. For $2^{nd}$- and
$3^{rd}$-generation leptoquarks, the cross sections for single LQ
production at $e^+e^-$ colliders are too small to be observable. These LQ's
can, however, be seen at $\gamma\gamma$ colliders. Depending on their
charges, $2^{nd}$-generation leptoquarks with masses between 700 and 900
GeV can be observed in $\gamma\gamma$ collisions at $\sqrt{s}=1$ TeV, while
at 500 GeV machines, LQ's whose mass is between roughly 300 and 400 GeV are
detectable. For $3^{rd}$-generation leptoquarks, the situation is not
nearly as promising. At $\sqrt{s}=500$ GeV, only LQ's with masses at most
190 GeV are observable, while at 1 TeV, it is possible to see LQ's with
$\ms$ up to just over 500 GeV. Thus, for $3^{rd}$-generation leptoquarks,
it seems that it is just as good, if not better, to look for signals from
pair production. Of course, if the LQ coupling strength were significantly
stronger than $\alpha_{em}$, as might be the case where the top quark is
involved, then single $3^{rd}$-generation LQ production would become more
promising.

\vskip 1truecm
\noindent
{\bf Acknowledgements}

G.B. would like to thank I.F. Ginzburg for helpful conversations. This work
was supported in part by the Natural Sciences and Engineering Research
Council of Canada, and by FCAR, Qu\'ebec.

\vfill\eject
\centerline{\bf References}
\vskip0.5truecm
\noindent
[1] I.F. Ginzburg, G.L. Kotkin, V.G. Serbo and V.I. Telnov, Pis'ma ZhETF
{\bf 34} (1981) 514; Sov.\ Yad.\ Fiz.\ {\bf 38} (1983) 372; Nucl.\ Instr.\
Methods {\bf 205} (1983) 47; I.F. Ginzburg, G.L. Kotkin, S.L. Panfil, V.G.
Serbo and V.I. Telnov, Sov.\ Yad.\ Fiz.\ {\bf 38} (1983) 1021; Nucl.\
Instr.\ Methods {\bf 219} (1984) 5; \hfil\break
[2] W. Buchm\"uller and D. Wyler, \plb{177}{86}{377}. \hfil\break
[3] M. Leurer, Weizmann Institute preprint WIS-93/26/March-Ph (1993).
\hfil\break
[4] H. Nadeau and D. London, \prd{47}{93}{3742}. \hfil\break
[5] J.L. Hewett and S. Pakvasa, \plb{227}{89}{178}. \hfil\break
[6] W. Buchm\"uller, R. R\"uckl and D. Wyler, \plb{191}{87}{442}.
\hfil\break
[7] I.F. Ginzburg, {\it Proc.\ IX Int.\ Workshop on Photon-Photon
Collisions}, eds. D. Caldwell and H. Paar (1993) 474.\hfil\break
[8] V.M. Budnev, I.F. Ginzburg, G.V. Meledin and V.G. Serbo,
\prep{15}{75}{182}. \hfil\break
[9] I.F. Ginzburg and V.G. Serbo, Novosibirsk preprint ``The
$\gamma\gamma\to Z l^+l^-$ and $\gamma\gamma\to Z q{\overline q}$ processes
at the polarized photon beams'' (unpublished).
\vfill\eject

\medskip
$$\vbox{\tabskip=0pt \offinterlineskip
\halign to \hsize{\strut#& \vrule#\tabskip 1em plus 2em minus .5em&
\hfil#\hfil &\vrule#& \hfil#\hfil &\vrule#& \hfil#\hfil &\vrule#&
\hfil#\hfil &\vrule#& \hfil#\hfil &\vrule#\tabskip=0pt\cr
\noalign{\hrule}
&& && && && && &\cr
&& $e^+e^- \to e^+ q S$: && \omit && $\qs$ && (a) $(\ms)_{max}$ &&
(b) $(\ms)_{max}$ &\cr
&& && && && && &\cr
\noalign{\hrule}
&& && && && && &\cr
&& \omit && $1^{st}$ gen. && $-1/3$, $-5/3$ && 475 && 960 &\cr
&& \omit && \omit && $-2/3$, $-4/3$ && 420 && 870 &\cr
&& && && && && &\cr
\noalign{\hrule}\noalign{\smallskip}\noalign{\hrule}
&& && && && && &\cr
&& $\gamma\gamma \to \ell^+ q S$: && \omit && $\qs$ &&
(a) $(\ms)_{max}$ && (b) $(\ms)_{max}$ &\cr
&& && && && && &\cr
\noalign{\hrule}
&& && && && && &\cr
&& \omit && $1^{st}$ gen. && $-1/3$, $-5/3$ && 480 && 970 &\cr
&& \omit && \omit && $-2/3$, $-4/3$ && 425 && 920 &\cr
&& && && && && &\cr
\noalign{\hrule}
&& && && && && &\cr
&& \omit && $2^{nd}$ gen. && $-1/3$ && 400 && 900 &\cr
&& \omit && \omit && $-5/3$ && 420 && 910 &\cr
&& \omit && \omit && $-2/3$ && 280 && 720 &\cr
&& \omit && \omit && $-4/3$ && 320 && 780 &\cr
&& && && && && &\cr
\noalign{\hrule}
&& && && && && &\cr
&& \omit && $3^{rd}$ gen. && $-1/3$ && --- && 180 &\cr
&& \omit && \omit && $-5/3$ && 140 && 530 &\cr
&& \omit && \omit && $-2/3$ && 100 && 400 &\cr
&& \omit && \omit && $-4/3$ && 190 && 540 &\cr
&& && && && && &\cr
\noalign{\hrule}  }}$$
\medskip
\noindent {Table 1: The largest LQ mass (in GeV) observable, for each of
the four LQ charges and for each of the three generations, in the processes
$e^+e^- \to e^+ q S$ and $\gamma\gamma \to \ell^+ q S$ at (a)
$\sqrt{s}=500$ GeV and (b) $\sqrt{s}=1$ TeV. $2^{nd}$- and
$3^{rd}$-generation LQ's cannot be seen at $e^+e^-$ colliders.}

\vfill\eject

\centerline{\bf Figure Captions}
\vskip0.5truecm
\noindent
1. The three sets of diagrams contributing to the process $e^+e^- \to e^+ q
S$. $q$ represents either a quark or an antiquark.

\vskip0.6truecm
\noindent
2. Diagrams contributing to the process $\gamma e^- \to q S$. $q$
represents either a quark or an antiquark.

\vskip0.6truecm
\noindent
3. Cross sections for single leptoquark production in $e^+e^-$ collisions
at (a) $\sqrt{s}=500$ GeV, (b) $\sqrt{s}=1$ TeV, for the 4 possible LQ
charges, $\qs = -1/3,-2/3,-4/3,-5/3$. The results are given for $k=1$. Here
we have used the nonzero quark mass as a regulator (see text).

\vskip0.6truecm
\noindent
4. Cross section for single leptoquark production in $e^+e^-$ collisions at
$\sqrt{s}=1$ TeV, using $\qs = -1/3$ and $k=2$ (this corresponds to $k=1$
in Ref.~\hewpak). Here we have used a 10 GeV $p_{\small{T}}$ cut as a
regulator (see text).

\vskip0.6truecm
\noindent
5. The four sets of diagrams contributing to the process $\gamma\gamma \to
\ell^+ q S$. $q$ represents either a quark or an antiquark.

\vskip0.6truecm
\noindent
6. Cross sections for single leptoquark production in $\gamma\gamma$
collisions for the 4 possible LQ charges, $\qs = -1/3,-2/3,-4/3,-5/3$, for
(a) $1^{st}$-generation LQ's at $\sqrt{s}=500$ GeV,
(b) $1^{st}$-generation LQ's at $\sqrt{s}=1$ TeV,
(c) $2^{nd}$-generation LQ's at $\sqrt{s}=500$ GeV,
(d) $2^{nd}$-generation LQ's at $\sqrt{s}=1$ TeV,
(e) $3^{rd}$-generation LQ's at $\sqrt{s}=500$ GeV,
(f) $3^{rd}$-generation LQ's at $\sqrt{s}=1$ TeV.
The results are given for $k=1$.

\bye